# Signatures of Majorana fermions in hybrid superconductor-semiconductor nanowire devices

V. Mourik[1]†, K. Zuo[1]†, S.M. Frolov[1], S.R. Plissard[2], E.P.A.M. Bakkers[1,2], L.P. Kouwenhoven[1]*

[1]*Kavli Institute of Nanoscience, Delft University of Technology,*
*2600 GA Delft, The Netherlands*
[2]*Department of Applied Physics, Eindhoven University of Technology,*
*5600 MB Eindhoven, The Netherlands*

*April 5, 2012*

Majorana fermions are particles identical to their own antiparticles. They have been theoretically predicted to exist in topological superconductors. We report electrical measurements on InSb nanowires contacted with one normal (Au) and one superconducting electrode (NbTiN). Gate voltages vary electron density and define a tunnel barrier between normal and superconducting contacts. In the presence of magnetic fields of order 100 mT we observe bound, mid-gap states at zero bias voltage. These bound states remain fixed to zero bias even when magnetic fields and gate voltages are changed over considerable ranges. Our observations support the hypothesis of Majorana fermions in nanowires coupled to superconductors.

†These authors contributed equally to this work.
*To whom correspondence should be addressed. E-mail: l.p.kouwenhoven@tudelft.nl

All elementary particles have an antiparticle of opposite charge (for example, an electron and a positron); the meeting of a particle with its antiparticle results in the annihilation of both. A special class of particles, called Majorana fermions, are predicted to exist that are identical to their own antiparticle *(1)*. They may appear naturally as elementary particles, or emerge as charge-neutral and zero-energy quasi-particles in a superconductor *(2, 3)*. Particularly interesting for the realization of qubits in quantum computing are pairs of localized Majoranas separated from each other by a superconducting region in a topological phase *(4-11)*.

Based on earlier semiconductor-based proposals, *(6)* and later *(7)*, Lutchyn et al. *(8)* and Oreg et al. *(9)* have outlined the necessary ingredients for engineering a nanowire device that should accommodate pairs of Majoranas. The starting point is a one-dimensional nanowire made of semiconducting material with strong spin-orbit interaction (Fig. 1A). In the presence of a magnetic field, $B$, along the axis of the nanowire (i.e. a Zeeman field), a gap is opened at the crossing between the two spin-orbit bands. If the Fermi energy, $\mu$, is inside this gap, the degeneracy is two-fold whereas outside the gap it is four-fold. The next ingredient is to connect the semiconducting nanowire to an ordinary s-wave superconductor (Fig. 1A). The proximity of the superconductor induces pairing in the nanowire between electron states of opposite momentum and opposite spins and induces a gap, $\Delta$. Combining this two-fold degeneracy with an induced gap creates a topological superconductor *(4-11)*. The condition for a topological phase is $E_Z > (\Delta^2 + \mu^2)^{1/2}$, with the Zeeman energy, $E_Z = g\mu_B B/2$ ($g$ is the Landé $g$-factor; $\mu_B$ the Bohr magneton). Near the ends of the wire, the electron density is reduced to zero and subsequently $\mu$ will drop below the subband energies such that $\mu^2$ becomes large. At the points in space where $E_Z = (\Delta^2 + \mu^2)^{1/2}$ Majoranas arise as zero-energy (i.e. mid-gap) bound states - one at each end of the wire *(4, 8-11)*.

Despite their zero charge and energy, Majoranas can be detected in electrical measurements. Tunneling spectroscopy from a normal conductor into the end of the wire should reveal a state at zero energy *(12-14)*. Here we report the observation of such zero-energy peaks and show that they rigidly stick to zero-energy while changing $B$ and gate voltages over large ranges. Furthermore, we show that this zero-bias peak is absent if we take out any of the necessary ingredients of the Majorana proposals, i.e. the rigid zero bias peak disappears for zero magnetic field, for a magnetic field parallel to the spin-orbit field, or when we take out the superconductivity.

We use InSb nanowires *(15)*, which are known to have strong spin-orbit interaction and a large $g$-factor *(16)*. From our earlier quantum dot experiments we extract a spin-orbit length $l_{so} \approx 200$ nm corresponding to a Rashba parameter $\alpha \approx 0.2$ eV·Å *(17)*. This translates to a spin-orbit energy scale $\alpha^2 m^* / (2\hbar^2) \approx 50$ μeV ($m^* = 0.015 m_e$ is the effective electron mass in InSb, $m_e$ is the bare electron mass). Importantly, the $g$-factor in bulk InSb is very large, $g \approx 50$, yielding $E_Z/B \approx 1.5$ meV/T. As shown below, we find an induced superconducting gap $\Delta \approx 250$ μeV. For $\mu = 0$ we thus expect to enter the topological phase for $B \sim 0.15$ T where $E_Z$ starts to exceed $\Delta$. The energy gap of the topological superconductor is estimated to be a few Kelvin *(17)*, if we assume a ballistic nanowire. The topological gap is significantly reduced in a disordered wire *(18, 19)*. We have measured mean free paths of ~300 nm in our wires *(15)*, implying a quasi-ballistic regime in micrometer long wires. With these numbers we expect Majorana zero-energy states to become observable below one Kelvin and around 0.15 T.

A typical sample is shown in Fig. 1B. We first fabricate a pattern of narrow (50 nm) and wider (300 nm) gates on a silicon substrate *(20)*. The gates are covered by a thin $Si_3N_4$ dielectric before we randomly deposit a low density of InSb nanowires. Next, we electrically contact those nanowires that have landed properly relative to the gates. The lower contact in Fig. 1B fully covers the bottom part of the nanowire. We have designed the upper contact to only cover half of the top part of the nanowire, avoiding complete screening of the underlying gates. This allows us to change the Fermi energy in the section of the nanowire (NW) with induced superconductivity. We have used either a normal (N) or superconducting (S) material for the lower and upper contacts, resulting in three sample variations: N-NW-S, N-NW-N and S-NW-S. Here we discuss our main results on the N-NW-S devices whereas the other two types, serving as control devices, are described in Ref. *(20)*.

To perform spectroscopy on the induced superconductor we create a tunnel barrier in the nanowire by applying a negative voltage to a narrow gate (dark green gate in Figure 1, B and C). A bias voltage applied externally between the N and S contacts drops almost completely across the tunnel barrier. In this setup the differential conductance *dI/dV* at voltage *V* is proportional to the density of states at energy $E = eV$, relative to the zero-energy, dashed line in Fig. 1C. Fig. 1D shows an example taken at $B = 0$. The two peaks at ±250 μeV correspond to the peaks in the quasi-particle density of states of the induced superconductor, providing a value for the induced gap, $\Delta \approx 250$ μeV. We generally find a finite *dI/dV* in between these gap edges. We observe pairs of resonances with energies symmetric around zero bias superimposed on non-resonant currents throughout the gap region. Symmetric resonances likely originate from Andreev bound states *(21, 22)*, whereas non-resonant current indicates that the proximity gap has not fully developed *(23)*.

Figure 2 summarizes our main result. Figure 2A shows a set of *dI/dV* versus *V* traces taken at increasing *B*-fields in 10 mT steps from zero (lowest trace) to 490 mT (top trace), offset for clarity. We again observe the gap edges at ±250 μeV. When we apply a *B*-field between ~100 and ~400 mT along the nanowire axis we observe a peak at $V = 0$. The peak has an amplitude up to ~0.05·$2e^2/h$ and is clearly discernible from the background conductance. Above ~400 mT we observe a pair of peaks. The color panel in Fig. 2B provides an overview of states and gaps in the plane of energy and *B*-field from -0.5 to 1 T. The observed symmetry around $B = 0$ is typical for all our data sets, demonstrating reproducibility and the absence of hysteresis. We indicate the gap edges with horizontal dashed lines (highlighted only for $B < 0$). A pair of resonances crosses zero energy at ~0.65 T with a slope of order $E_Z$ (highlighted by dotted lines). We have followed these resonances up to high bias voltages in *(20)* and identified them as Andreev states bound within the gap of the bulk, NbTiN superconducting electrodes (~2 meV). By contrast, the zero-bias peak sticks to zero energy over a range of $\Delta B \sim 300$ mT centered around ~250 mT. Again at ~400 mT we observe two peaks located at symmetric, finite biases.

In order to identify the origin of these zero-bias peaks (ZBP) we need to consider various options, including the Kondo effect, Andreev bound states, weak antilocalization and reflectionless tunneling, versus a conjecture of Majorana bound states. ZBPs due to the Kondo effect *(24)* or Andreev states bound to s-wave superconductors *(25)* can occur at finite *B*. However, when changing *B* these peaks then split and move to finite energy. A Kondo resonance moves with twice $E_z$ *(24)*, which is easy to dismiss as the origin for our zero-bias peak because of the large *g*-factor in InSb. (Note that even a Kondo effect from an impurity with *g* = 2 would be discernible.) Reflectionless tunneling is an enhancement of Andreev reflection by time-reversed paths in a diffusive normal region *(26)*. As in the case of weak antilocalization, the resulting ZBP is maximal at *B* = 0 and disappears when *B* is increased, see also *(20)*. We thus conclude that the above options for a ZBP do not provide natural explanations for our observations. We are not aware of any mechanism that could explain our observations, besides the conjecture of a Majorana.

To further investigate the zero-biasness of our peak, we measure gate voltage dependences. Fig. 3A shows a color panel with voltage sweeps on gate 2. The main observation is the occurrence of two opposite types of behavior. First, we observe peaks in the density of states that change with energy when changing gate voltage (e.g. highlighted with dotted lines), these are the same resonances as shown in Fig. 2B and analyzed in *(20)*. The second observation is that the ZBP from Fig. 2, which we take at 175 mT, remains stuck to zero bias while changing the gate voltage over a range of several volts. Clearly, our gates work since they change the Andreev bound states by ~0.2 meV per Volt on the gate. Panels (B) and (C) underscore this observation with voltage sweeps on a different gate, number 4. (B) shows that at zero magnetic field no ZBP is observed. At 200 mT the ZBP becomes again visible in (C). Comparing the effect of gates 2 and 4, we observe that neither moves the ZBP away from zero.

Initially, Majorana fermions were predicted in single-subband, one-dimensional wires *(8, 9)*, but further work extended these predictions to multi-subband wires *(27-30)*. In the nanowire section that is uncovered we can gate tune the number of occupied subbands from 0 to ~4 with subband separations of several meV. Gate tuning in the nanowire section covered with superconductor is much less effective due to efficient screening. The number of occupied subbands in this part is unknown, but it is most likely multi-subband. As shown in Figures S9 and S11 of *(20)* we do have to tune gate 1 and the tunnel barrier to the right regime in order to observe the ZBP.

We have measured in total several hundred panels sweeping various gates on different devices. Our main observations *(20)* are (1) ZBP exists over a substantial voltage range for every gate starting from the barrier gate until gate 4, (2) we can occasionally split the ZBP in two peaks located symmetrically around zero, and (3) we can never move the peak away from zero to finite bias. Data sets such as those in Figures 2 and 3 demonstrate that the ZBP remains stuck to zero energy over considerable changes in *B* and gate voltage $V_g$.

Figure 3D shows the temperature dependence of the ZBP. We find that the peak disappears at around ~300 mK, providing a thermal energy scale of $k_BT \sim 30$ μeV. The full-width at half-maximum at the lowest temperature is ~ 20 μeV, which we believe is a consequence of thermal broadening as $3.5 \cdot k_BT(60$ mK$) = 18$ μeV.

Next we verify explicitly that all the required ingredients in the theoretical Majorana proposals (Figure 1A) are indeed essential for observing the ZBP. We have already verified that a non-zero *B*-field is needed. Now, we test if spin-orbit interaction is crucial for the absence or presence of the ZBP. Theory requires that the external *B* has a component perpendicular to $B_{so}$. We have measured a second device in a different setup containing a 3D vector magnet such that we can sweep the *B* field in arbitrary directions. In Figure 4 we show *dI/dV* versus *V* while varying the angle for a constant field magnitude. In Figure 4A the plane of rotation is approximately equal to the plane of the substrate. We clearly observe that the ZBP comes and goes with angle. The ZBP is completely absent around π/2, which thereby we deduce as the direction of $B_{so}$. In Figure 4B the plane of rotation is perpendicular to $B_{so}$. Indeed we observe that the ZBP is now present for all angles, because *B* is now always perpendicular to $B_{so}$. These observations are in full agreement with expectations for the spin-orbit direction in our samples *(17, 31)*. We have further verified that this angle dependence is not a result of the specific magnitude of *B* or a variation in *g*-factor *(20)*.

As a last check we have fabricated and measured a device of identical design but with the superconductor replaced by a normal Au contact (i.e. a N-NW-N geometry). In this sample we have not found any signature of a peak that sticks to zero bias while changing both *B* and $V_g$ *(20)*. This test experiment shows that superconductivity is indeed also an essential ingredient for our ZBP.

To summarize, we have reproduced in three different devices and in two different setups our observation of a rigid ZBP. Our general observations are: (1) a ZBP appears at finite *B* and sticks to zero bias over a range from 0.07 to 1 T, (2) the ZBP remains at zero bias while changing the voltage on any of our gates over significant ranges, (3) the ZBP comes and goes with the angle of the *B* field with respect to the wire axis, which is in agreement with the expected spin-orbit interaction, (4) the rigid ZBP is absent when the superconductor is replaced by a normal conductor. Based on these observations we conclude that our spectroscopy experiment provides evidence for the existence of Majorana fermions.

Improving the electron mobility and optimizing the gate coupling will enable mapping out the phase diagram of the topological superconductor in the plane of $E_Z$ and *μ (27-30)*. It will be interesting to control the subband occupation underneath the superconductor down to a single subband in order to make direct comparisons to theoretical models. Currently, we probe induced gaps and states from all occupied subbands, each with a different coupling to the tunnel barrier. The topological state in the topmost subband likely has the weakest coupling to the tunnel barrier. Single subband models *(8, 9)* predict that one should observe a closing of the topological gap, however, in multi-subband systems this gap closing may not be visible. The constant gap in Fig. 2 may come from lower subbands. The presence of multi-subbands together with our finite temperature may also be the reason that our ZBP is currently only ~5% of the theoretical zero-temperature limit of $2e^2/h$ *(12, 14)*.

Finally, we note that this work does not address the topological properties of Majorana fermions. The first step towards demonstrating topological protection would be the observation of conductance quantization *(12, 32)*. Second, in a Josephson tunnel junction with phase difference $\varphi$ and with a pair of Majoranas on either side, the current-phase relation becomes proportional to $\sin(\varphi/2)$. The factor 2 is another distinct Majorana signature, which should be observable as an $h/e$ flux periodicity in a SQUID measurement *(8, 9)*. The last type of experiment involves the exchange of Majoranas around each other. Such braiding experiments can reveal their non-Abelian statistics, which is the ultimate proof of topologically protected Majorana fermions *(33-35)*.

We thank David Thoen and Teun Klapwijk for sharing their NbTiN technology. For discussions and assistance we thank Anton Akhmerov, Jason Alicea, Carlo Beenakker, Michael Freedman, Fabian Hassler, George Immink, Han Keijzers, Charles Marcus, Stevan Nadj-Perge, Yuli Nazarov, Ilse van Weperen, Michael Wimmer, David van Woerkom. This work has been supported by ERC, NWO/FOM Netherlands Organization for Scientific Research and Microsoft Corporation Station Q.

**Figure Captions**

**Fig. 1. (A)** Outline of theoretical proposals. **Top:** Conceptual device layout with a semiconducting nanowire in proximity to an s-wave superconductor. An external *B*-field is aligned parallel to the wire. The Rashba spin-orbit interaction is indicated as an effective magnetic field, $B_{so}$, pointing perpendicular to the nanowire. The red stars indicate the expected locations of a Majorana pair. **Bottom:** Energy, *E*, versus momentum, *k*, for a 1D wire with Rashba spin-orbit interaction, which shifts the spin-down band (blue) to the left and spin-up band (red) to the right. Blue and red parabola are for *B* = 0. Black curves are for *B* ≠ 0, illustrating the formation of a gap near *k* = 0 of size $g\mu_B B$. ($\mu$ is the Fermi energy with $\mu$ = 0 defined at crossing of parabolas at *k* = 0). The superconductor induces pairing between states of opposite momentum and opposite spin creating a gap of size Δ. **(B)** Implemented version of theoretical proposals. Scanning electron microscope image of the device with normal (N) and superconducting (S) contacts. The S-contact only covers the right part of the nanowire. The underlying gates, numbered 1 to 4, are covered with a dielectric. (Note that gate 1 connects two gates and gate 4 connects four narrow gates; see **(C)**.) **(C) Top:** Schematic of our device. **Down:** illustration of energy states. Green indicates the tunnel barrier separating the normal part of the nanowire on the left from the wire section with induced superconducting gap, Δ. (In **(B)** the barrier gate is also marked green.) An external voltage, *V*, applied between N and S drops across the tunnel barrier. Red stars again indicate the idealized locations of the Majorana pair. Only the left Majorana is probed in this experiment. **(D)** Example of differential conductance, *dI/dV*, versus *V* at *B* =0 and 65 mK, serving as a spectroscopic measurement on the density of states in the nanowire region below the superconductor. Data from device 1. The two large peaks, separated by 2Δ, correspond to the quasi-particle singularities above the induced gap. Two smaller subgap peaks, indicated by arrows, likely correspond to Andreev bound states located symmetrically around zero energy. Measurements are performed in dilution refrigerators using standard low-frequency lock-in technique (frequency 77 Hz, excitation 3 μV) in the four-terminal (devices 1 and 3) or two-terminal (device 2) current-voltage geometry.

**Fig. 2.** Magnetic field dependent spectroscopy. **(A)** *dI/dV* versus *V* at 70 mK taken at different *B*-fields (from 0 to 490 mT in 10 mT steps; traces are offset for clarity, except for the lowest trace at *B* = 0). Data from device 1. **(B)** Color scale plot of *dI/dV* versus *V* and *B*. The zero-bias peak is highlighted by a dashed oval. Dashed lines indicate the gap

edges. At ~0.6 T a non-Majorana state is crossing zero bias with a slope equal to ~3 meV/T (indicated by sloped dotted lines). Traces in (**A**) are extracted from (**B**).

**Fig. 3.** Gate voltage dependence. (**A**) 2D color plot of *dI/dV* versus *V* and voltage on gate 2 at 175 mT and 60 mK. Andreev bound states cross through zero bias, for example near -5 V (dotted lines). The ZBP is visible from -10 to ~5 V (although in this color setting it is not equally visible everywhere). Split peaks are observed in the range of 7.5 to 10 V *(20)*. In (**B**) and (**C**) we compare voltage sweeps on gate 4 for 0 and 200 mT with the zero bias peak absent and present, respectively. Temperature is 50 mK. (Note that in (**C**) the peak extends all the way to -10 V *(19)* (**D**) Temperature dependence. *dI/dV* versus *V* at 150 mT. Traces have an offset for clarity (except for the lowest trace). Traces are taken at different temperatures (from bottom to top: 60, 100, 125, 150, 175, 200, 225, 250 and 300 mK). *dI/dV* outside ZBP at *V* = 100 μeV is 0.12±0.01·$2e^2/h$ for all temperatures. A full-width at half-maximum of 20 μeV is measured between arrows. All data in this figure are from device 1.

**Fig. 4.** Magnetic field orientation dependence. *dI/dV* versus *V* and varying the angle of *B* at fixed magnitude. Data from device 2 measured in a different setup at ~150 mK. Zero angle is along the nanowire for both panels. (**A**) Rotation of |*B*| = 200 mT in the plane of the substrate. The ZBP is maximum when *B* is parallel and absent when *B* is perpendicular to the wire. (**B**) Rotation of |*B*| = 150 mT in the plane perpendicular to $B_{so}$. The ZBP is now present for all angles. The panels on top show linecuts at angles with corresponding colors in (**A**) and (**B**). Panels on the right side illustrate from top to bottom: (1) For *B* perpendicular to $B_{so}$ a gap opens lifting fermion doubling, as is required for Majoranas, (2) For *B* parallel to $B_{so}$ the two spin bands from Fig. 1A shift vertically by $2E_Z$. In this configuration a zero-energy Majorana is not expected, (3) panel of rotation of *B* for data in (**A**), (4) panel of rotation of *B* for data in (**B**).

Figure 1

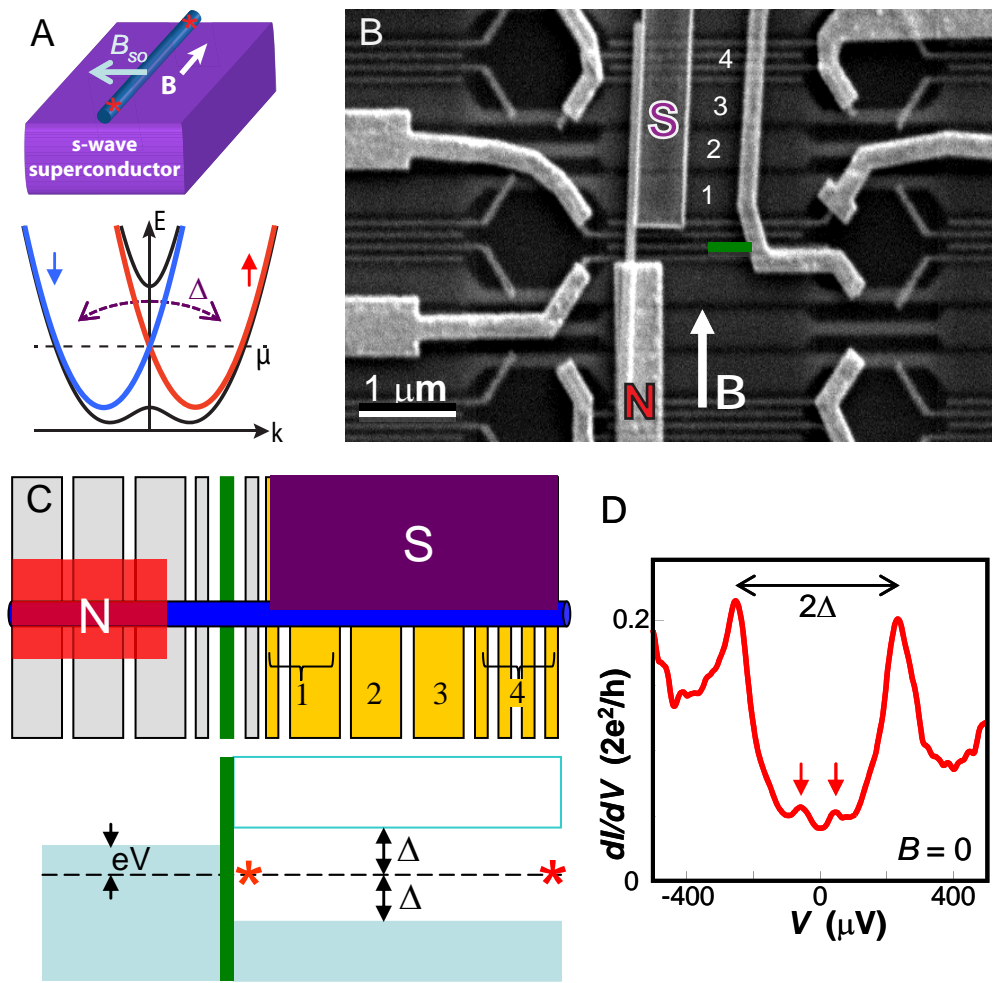

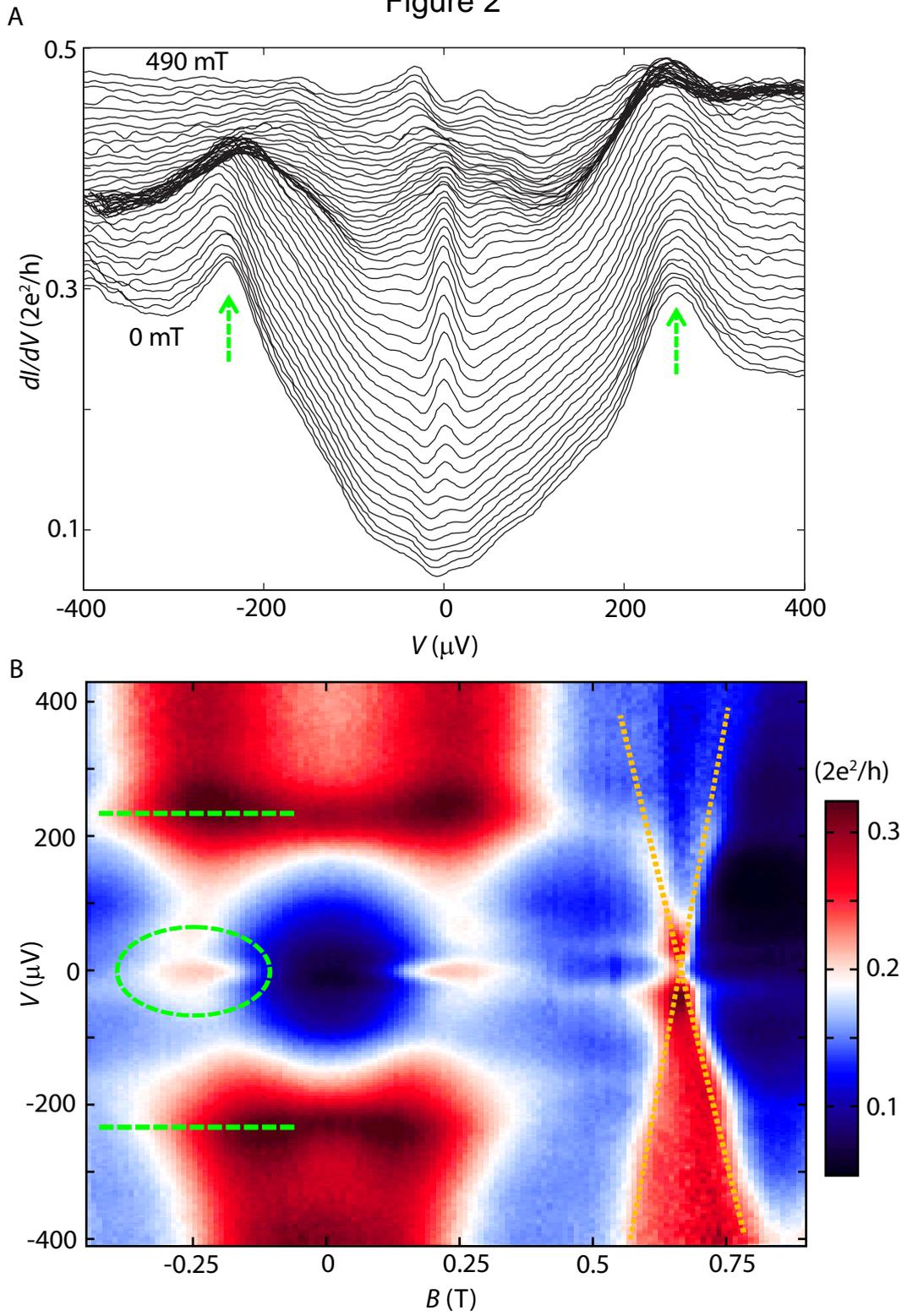

Figure 3

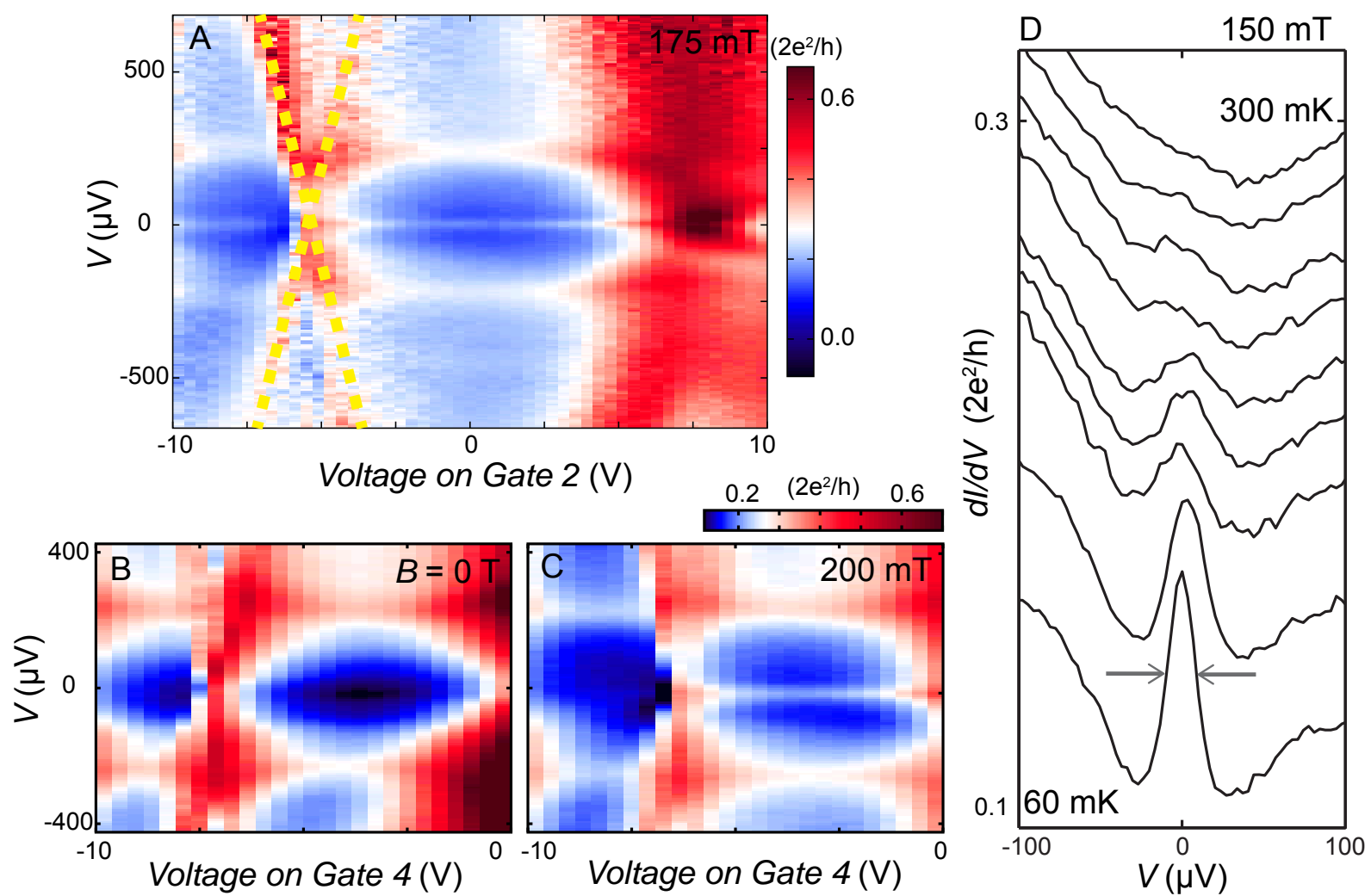

# Figure 4

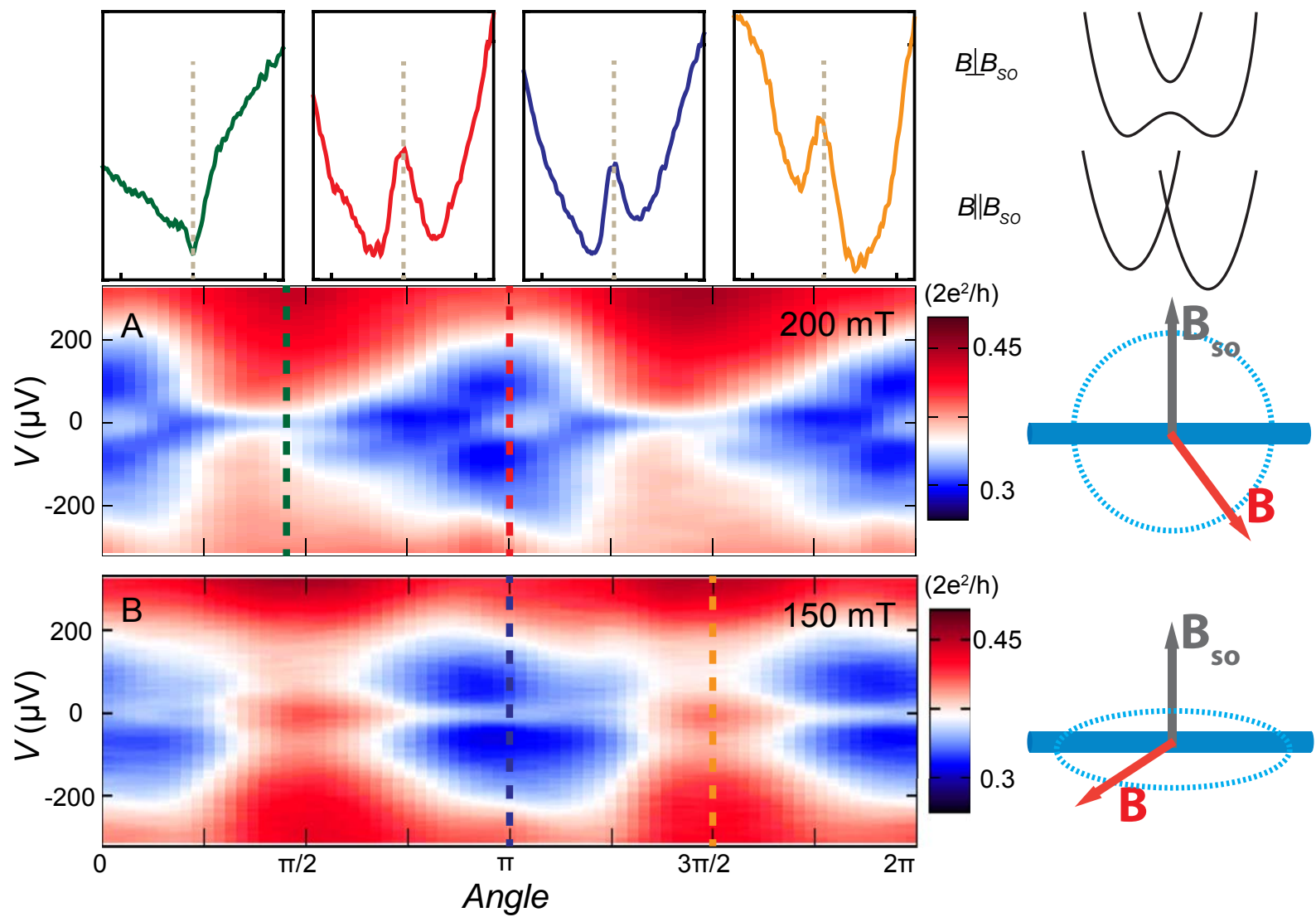

Supplementary Materials:
# Signatures of Majorana fermions in hybrid superconductor-semiconductor nanowire devices

V. Mourik, K. Zuo, S.M. Frolov, S.R. Plissard, E.P.A.M. Bakkers and L.P. Kouwenhoven

**Summary of Supplementary Figures**

Below we provide details of nanowire sample fabrication as well as supporting measurements from multiple devices. The main observations of the paper, i.e. zero-bias peak (ZBP) that appears at finite magnetic field and persists over a significant range of field and gate voltages, are reproduced in three N-NW-S devices measured in two setups (Figs. S1, S3, S6,S7, S10, S11). Furthermore we demonstrate S-NW-S devices and N-NW-N devices (Figs. S12-S14). In S-NW-S devices persistent zero-bias peaks are also observed, however they cannot be distinguished from Josephson supercurrents. In N-NW-N devices zero-bias peaks are also observed for a small range of gate voltages (Fig. S13, S14), however only when gate- and field-tunable states pass through zero bias. This indicates that superconductivity is a required ingredient for the observation of a persistent zero-bias peak.

Specifically, we present more examples of magnetic field dependences of the zero bias peak in N-NW-S devices (Figs. S3, S6, S7, S10). These data establish the magnetic field range of the zero bias peak from ~70 mT and up to 1.0 T (varying for different gate settings). Additional gate dependences investigate the splitting of the zero-bias peak (Figs. S4, S5). Examples of tunnel barrier gate dependences are provided in Figs. S9 and S11. Finally, other features that occur at zero bias are studied in Figs. S2 and S8. In Fig. S2 we identify Andreev bound states confined in the nanowire segment covered by the superconductor. In Fig. S8 we investigate zero-bias peaks observed at zero magnetic field.

**Author contributions.** L.P.K. and S.M.F. supervised the experiments; S.R.P. and E.P.A.M.B. grew the nanowires; V.M. and K.Z. fabricated nanowire devices; V.M., K.Z. and S.M.F. performed the measurements; V.M., K.Z., S.M.F. and L.P.K. analyzed the data; the manuscript has been prepared with contributions from all authors.

## Figure S1: N-NW-S device fabrication

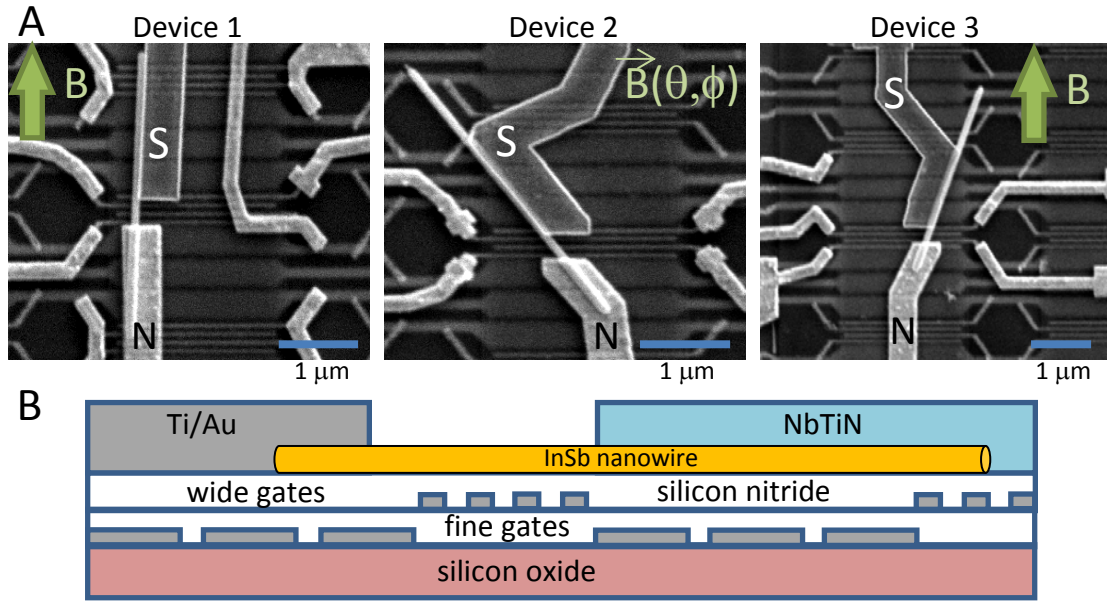

**A**, SEM images of three N-NW-S (Normal-Nanowire-Superconductor) devices in which the main findings of this paper are reproduced. Field directions are indicated with arrows. Device 2 was measured in a 3-axis vector magnet. Devices are fabricated simultaneously. Nanowire diameters are 110±10 nm (devices 1 and 3) and 100±10 nm (device 2). **B**, Schematic of a device cross-section.

**Nanowire growth details.** InSb nanowires are grown by metalorganic vapor phase epitaxy from gold catalysts, as described in Ref. (*15*). The wires in this work are grown on Si substrates. First, stems that consists of InP and InAs segments are grown. Then a stacking-fault and dislocation-free zincblende InSb segment of high mobility ($10^4$-$5 \cdot 10^4$ cm$^2$/(Vs)) is grown in the 111 crystal direction. A single batch of wires is used for all N-NW-S devices in this paper.

**N-NW-S device fabrication procedure**
1) p-doped silicon substrates are covered by 285 nm of thermal oxide. Due to screening from local gates substrates are ineffective as back gates.
2) A periodic pattern of 15 micron long and 300 nm wide Ti/Au gates (5 nm/10nm) is defined by 100 kV electron beam lithography and electron beam evaporation.
3) Bottom gate layer is covered by 40 nm of lithographically defined and d.c. sputtered $Si_3N_4$ dielectric. Areas for contacts to gates are left free of dielectric.
4) A second layer of finer gates (50 nm wide, 50 nm spacing) is defined using the same method. Fine gates are fabricated in a separate step to reduce proximity exposure.
5) A second layer of $Si_3N_4$ covers both fine and wide gates. Thus, wide gates are covered by 80 nm of dielectric, fine gates are covered by 40 nm of dielectric.
6) InSb nanowires of 80-120 nm diameter are transferred onto the substrate containing gate patterns. Nanowires land randomly, some are selected for contacting.
7) Superconducting contacts are defined by sputtering NbTiN (75 nm) from a Nb/Ti target (70/30 at. %) with thin film critical temperature $T_c$ ~ 7 K. Sputtering done in the group of T.M. Klapwijk with assistance of D.J. Thoen. A window in the 200 nm thick PMMA 950k resist has a boundary along the center of the nanowire with alignment accuracy of 20-30 nm. Prior to sputtering nanowires are etched in Argon plasma.
8) Normal Ti/Au contacts (20 nm Ti/125 nm Au) are made to the nanowires and to the gates. Prior to the deposition of Ti/Au the nanowires are passivated in ammonium sulfide.

Figure S2: Large bias scans to identify Andreev bound states

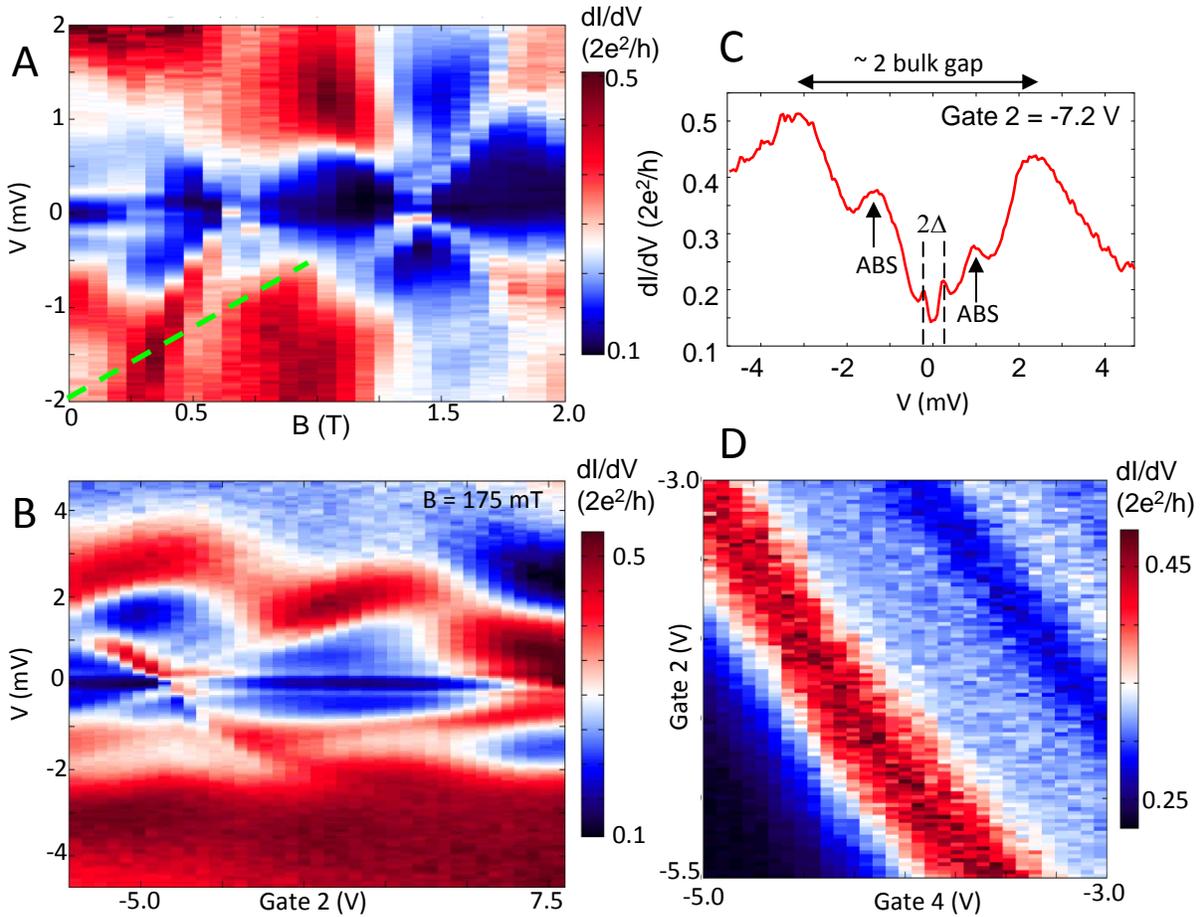

In this figure we investigate the states that cross zero bias and appear in Figs. 2 and 3 of the main paper. **A**, Magnetic field dependence of dI/dV extended to higher source-drain voltages. Despite low resolution the induced (soft) gap is observed at ~0.25 mV, and the zero-bias peak is visible between 200 and 700 mT. Two pairs of states exhibit a strong magnetic field dependence, and cross zero at ~0.7 T and ~1.4 T. **Notably, these states extend above the induced gap, but are also present within this gap.** Dashed line indicates the Zeeman energy ½$g\mu_B B$ for g=50 (the bulk value in InSb). The larger slope of the observed states can be due to field expulsion from the superconductor. **B**, States that cross zero bias are also tunable with gates 2,3,4 (gate 2 dependence shown). In this scan over a larger range of V they are traceable to the source-drain voltage of ~3 mV, which is on the scale of the bulk gap in the NbTiN electrode. These plots are reminiscent of numerical data by C. Bena *(36)*. **We interpret these states as Andreev Bound States (ABS) confined between the bulk superconductor and the gate-defined tunnel barrier.** As expected for ABS, these states come in pairs, one at positive and one at negative bias. **C**, Linecut from **B** showing the induced gap at 0.25 mV, a pair of ABS resonances near 1.5 mV and an enhanced conductance on the scale of the gap of NbTiN above 2 mV. **D**, A plot of dI/dV at V = 640 μV, B = 175 mT showing that the same ABS resonance (red) can be tuned by two gates underneath the superconductor that are 400 nm apart. Apparently the ABS are extended over the entire segment of the nanowire that is underneath the superconductor, suggesting a finite density of states within the apparent gap even deep underneath the superconductor. ABS increase conductance much stronger than ZBP. This may come about if ABS belong to lower subbands and/or have a stronger penetration into the tunnel barrier. (Data from N-NW-S Device 1, T = 60 mK)

## Figure S3: Additional Magnetic Field Dependences

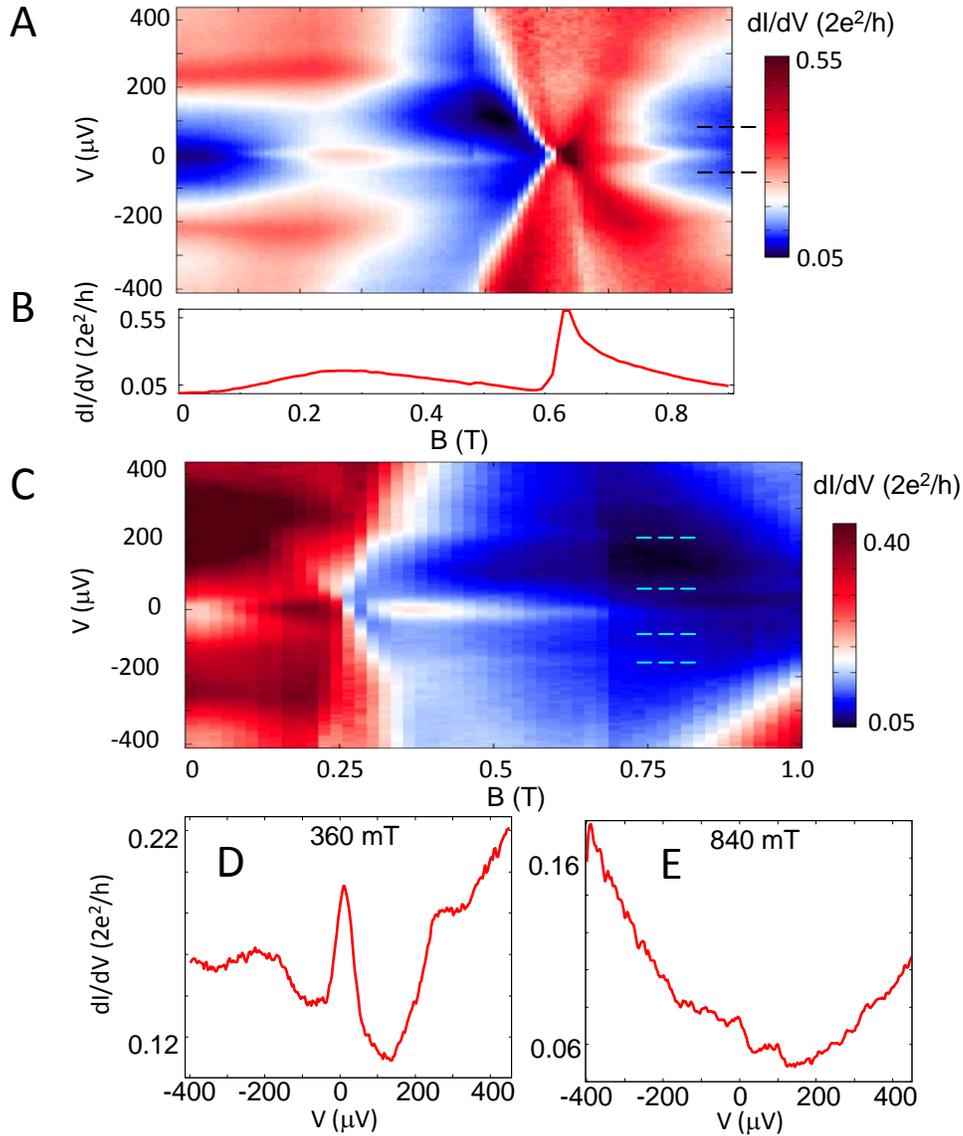

In this figure we show more examples of magnetic field dependences from device 1, to complement data in Fig.2 of the main text. **A**, Magnetic field dependence of dI/dV. Zero-bias peak is shown extending up to ~0.9 T. In the vicinity of B = 0.6 T a pair of ABS cross zero bias (Gate 1 = -0.325 V, Gate 2 = 0.2 V, Gate 3 = -1.6 V, Gate 4 = -4.0 V). **B**, Conductance at zero bias is suppressed at fields immediately below the ABS crossing point and enhanced at the crossing. The same behavior is observed at finite bias. The asymmetric shape of the dI/dV trace is reminiscent of a Fano resonance. We speculate that a Fano resonance results from interference between an ABS and a continuum of states within the bulk gap. We observe that the height of the ZBP is strongly influenced when an ABS crosses zero. The ZBP itself seems to persist throughout an ABS crossing. **C**, After re-tuning Gate 2 the ABS crossing point is shifted to lower magnetic field B~0.2 T (Gate 1 = -0.325 V, Gate 2 = -3.7 V, Gate 3 = -1.6 V, Gate 4 = -4.0 V). The zero-bias peak is observed starting from B = 0.1 T. The ZBP is traceable to B = 1.0 T in color scale. However the amplitude of the ZBP drops for B > 0.7 T. A number of resonances that run parallel to ZBP, i.e. that do not have a magnetic field dependence, are visible within the induced gap (dashed lines in panels **A** and **C**). **D** and **E**, Linecuts from panel **C**. (Data from N-NW-S Device 1, T = 60 mK)

## Figure S4: Gate 4 scans at different magnetic fields

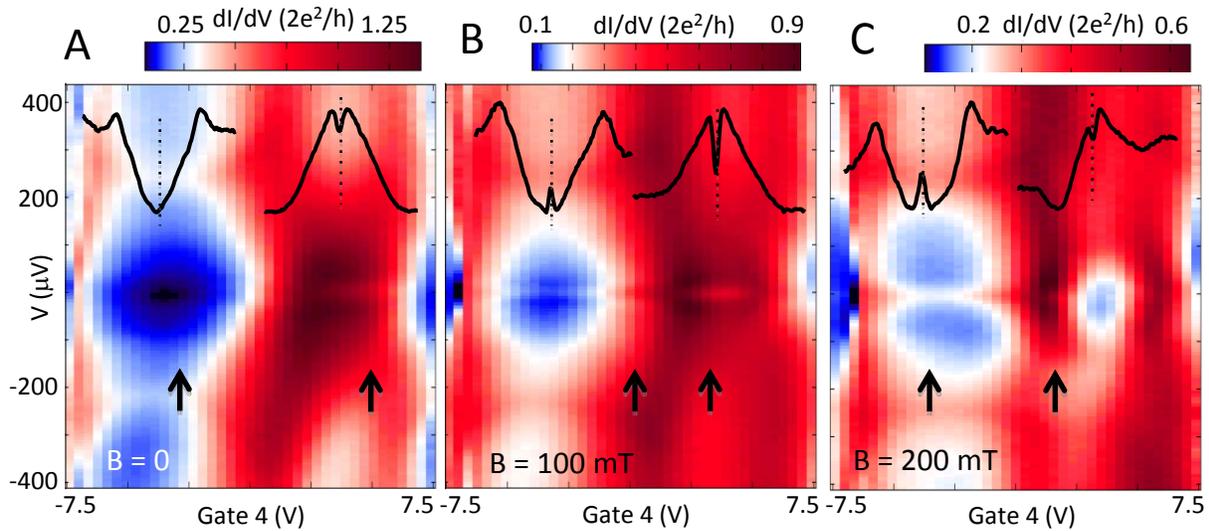

**A-C**, Gate 4 voltage dependences of dI/dV at three values of magnetic field. These data are an extension of a set displayed in Figs. 3B, 3C of the main paper. Zero-bias peaks appear at finite magnetic field where they are best visible in the low conductance regions (blue regions), which are not obscured by ABS resonances (ABS appear as red regions in the color scale). In all panels, including at $B = 0$, we observe two peaks in the region of high conductance (see line cuts on the right in each panel). Taken at face value, the data in this figure does not suggest a connection between the ZBP and the split peak. However, currently we do not have a precise understanding of the various split peaks and their relation to the rigid ZBP. In each panel two linecuts illustrate dI/dV behavior at gate settings marked by arrows. Dashed lines indicate zero bias. (Data from N-NW-S Device 1, Gate 1 = -0.325 V, Gate 2 = 0.2 V, Gate 3 = -1.6 V, T = 60 mK)

## Figure S5: Apparent Splitting of Zero-Bias Peak

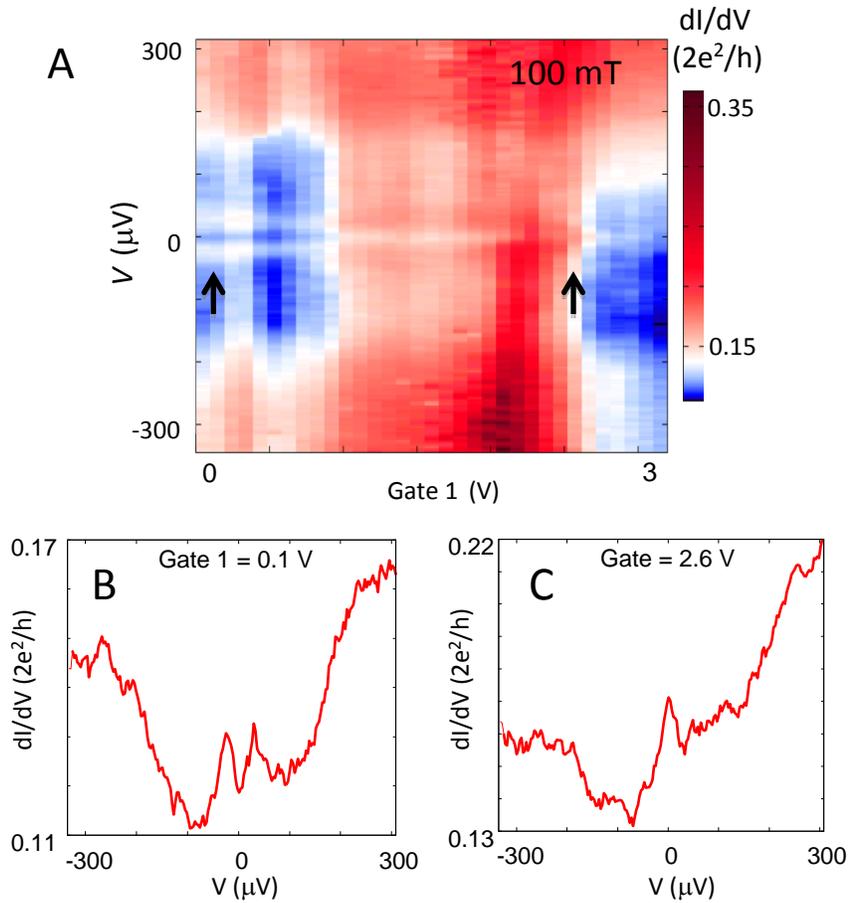

**A**, In this figure we show that conductance near zero bias can be tuned from a single peak at zero bias to a pair of narrow split peaks also in the regime of low conductance, away from ABS resonances. Black arrows indicate traces displayed in panels **B** and **C**. In the context of Majorana fermions split zero-bias peaks can be understood as coupling of two nearby Majoranas. However, split peaks in the low conductance regime (below 0.3·2e²/h) and at low magnetic fields (100-300 mT) are a relatively rare observation in our current experiment. They are not observed frequently enough to draw conclusions in the context of overlapping Majoranas. A detailed investigation of split peaks is beyond the scope of the present manuscript. A device with optimized gate geometry might provide a better setting to investigate this effect. (Data from N-NW-S Device 3.)

## Figure S6: In-plane field rotation data

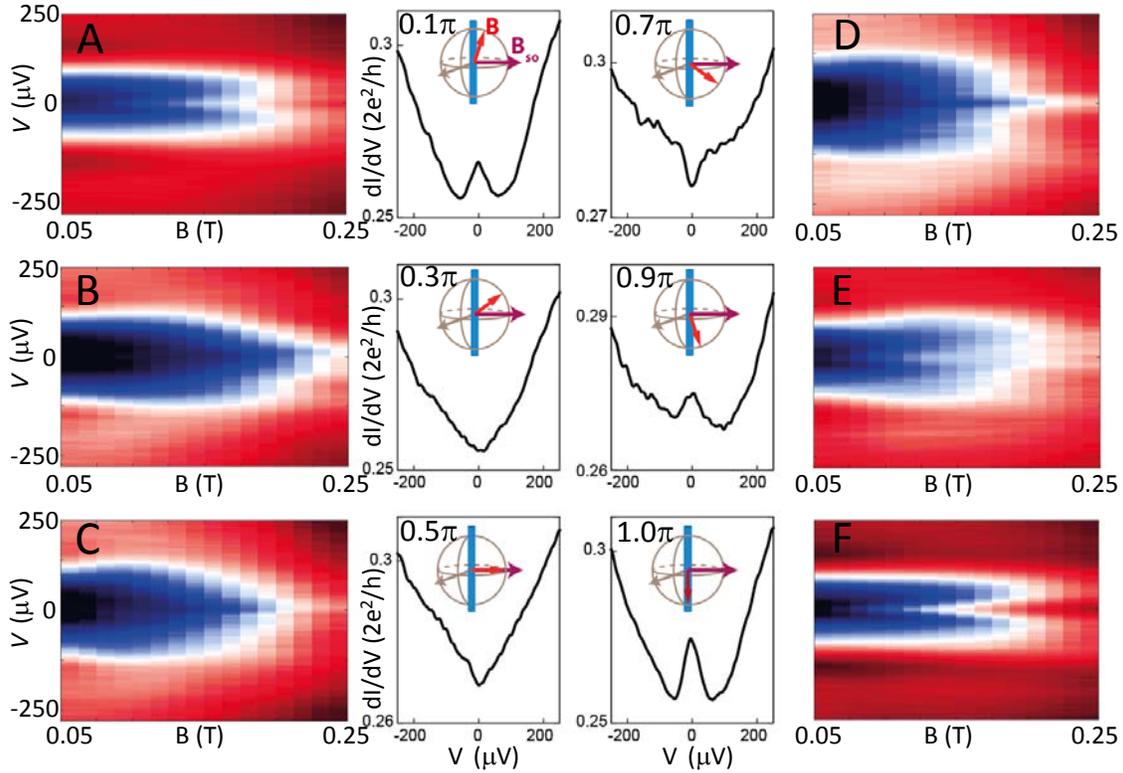

This figure shows data complementary to Fig. 4 of the main paper. It also demonstrates that field dependences of the ZBP obtained from N-NW-S device 2 are in qualitative agreement with those from device 1. **A-F**, Magnetic field vs. bias maps of dI/dV. For each panel magnetic field is applied at a different angle in the plane of the substrate (accuracy 10 degrees). Inner panels show traces at B = 143 mT. Insets to the inner panels illustrate the orientation of the magnetic field for each panel (red arrow), blue is the nanowire axis, purple is the spin-orbit field direction. **The zero-bias peak disappears when the magnetic field is perpendicular to the nanowire. This was determined as the direction of the spin-orbit field in previous work on the same nanowires** *(17)*. The modulation of ZBP amplitude is observed in the range 0.1T – 0.3T. This demonstrates that the disappearance of the peak is not due to a variation in the onset field of the ZBP induced by g-factor anisotropy. (Data from N-NW-S Device 2, T ~ 150 mK)

## Figure S7: Out-of-plane field rotation data

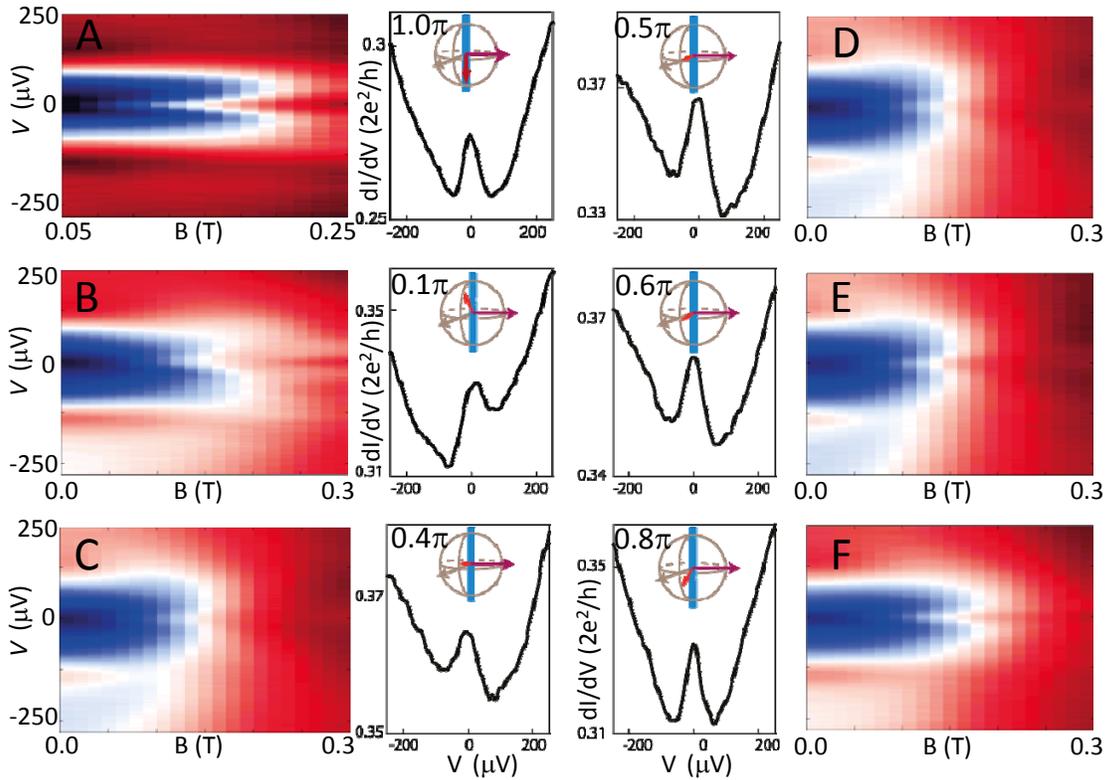

This figure shows data complementary to Fig. 4 of the main paper. **A-F**, Magnetic field vs. bias maps of dI/dV. For each panel magnetic field is applied at a different angle in the plane perpendicular to the spin-orbit field $B_{SO}$. Inner panels show traces at B = 150 mT. Insets to the inner panels illustrate the orientation of the magnetic field for each panel (red arrow), blue is the nanowire axis, purple is the spin-orbit field direction. **Zero bias peaks of similar amplitude are observed for all orientations perpendicular to spin-orbit field.** (Data from N-NW-S Device 2, T ~ 150mK.) Note that panel S7 **A** is identical to panel S6 **F**.

## Figure S8: Zero-bias peak at zero magnetic field

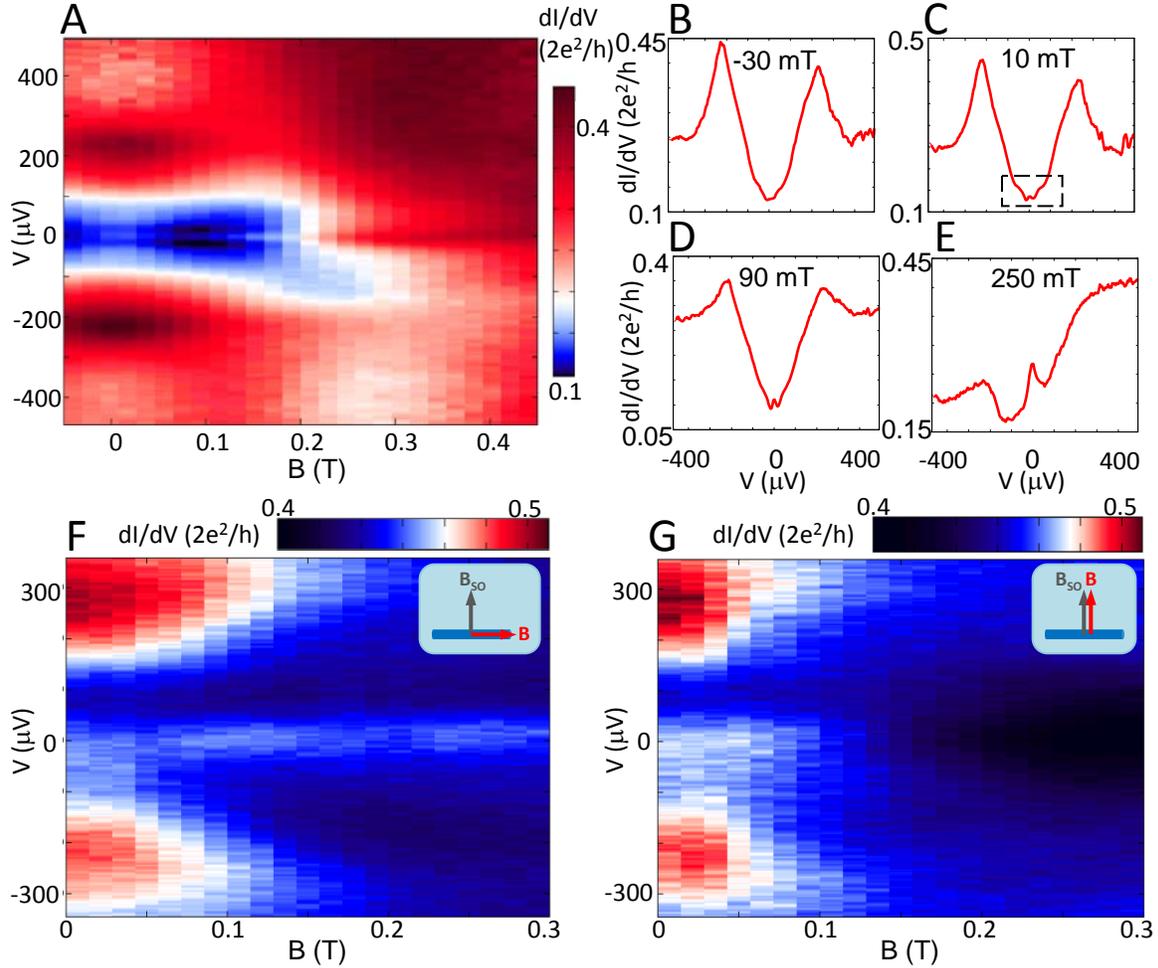

**A**, A small, but discernible zero bias peak is sometimes observed also for B ≈ 0. This peak is observed for unique gate settings (gate F3 = -0.14 V, gate F2 = -0.1875 V, gate 1 = 0.54 V, gate 2 = 4.665 V, gate 3 = -1.6 V, gate 4 = -4 V). The ZBP at zero field is observed much less often than the robust ZBP at finite magnetic field. (Data in **A-E** are from N-NW-S Device 1). **B-E**, Linecuts from panel **A** at different magnetic fields. The zero-bias feature in the vicinity of B = 0 has a height of 0.005·$2e^2/h$ (dashed box, panel **C**). **F** and **G**, The zero-bias peak at zero magnetic field is reproduced in N-NW-S device 2 for certain gates voltage settings combination. However, when the magnetic field is aligned with the spin-orbit field the zero-bias peak is suppressed starting at ~100 mT, the typical onset for the finite-field ZBP. (**F**: field along the wire, **G**: field along the spin-orbit field, perpendicular to the wire).

Possible origins of ZBP at zero field include weak antilocalization, reflectionless tunneling and Josephson effect. Supercurrent flow is unlikely in our N-NW-S devices, since the critical field of Ti (part of Ti/Au normal contact) does not exceed 30 mT, and superconductivity in Ti is further weakened by the inverse proximity effect from a thick gold layer. The field scale for both weak antilocalization and reflectionless tunneling is determined by $B_0 \sim (h/e)/(A)$, where A is a characteristic area of an electron trajectory perpendicular to the field direction. While field expulsion from the superconductor complicates the prediction of $B_0$, it can be estimated in the 100's of milliTesla range.

Figure S9: Pinch-off gate traces of barrier gate

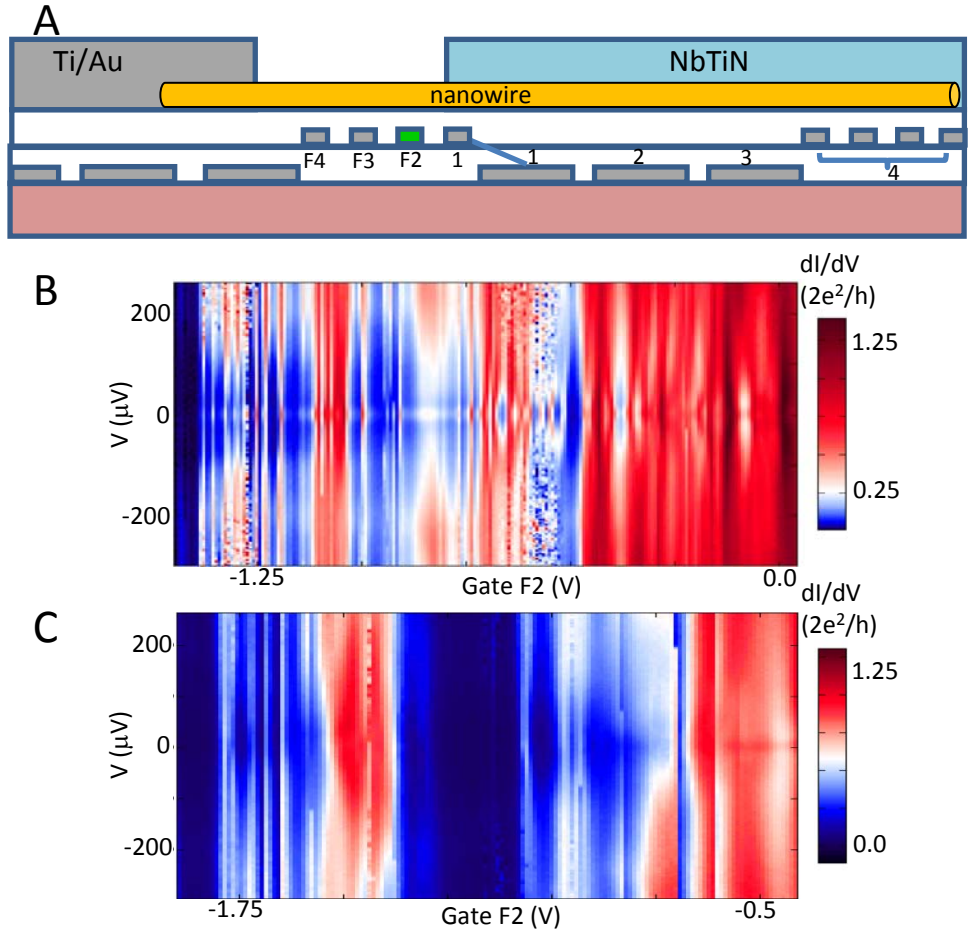

Fig. 3 in the main paper shows the effect of the gates underneath the superconductor. Here we present the effect of a tunnel barrier gate, the so-called pinch-off traces. **A**, Device 1 schematic with bottom gates labeled. Wide gate 1 is connected to an adjacent fine gate. Gate 4 consists of four narrow gates. Details of gate layout in the other two N-NW-S devices differ. **B**, Conductance map obtained by sweeping the barrier gate F2 from open regime (near 0 V) towards pinch-off at negative voltages. A zero-bias peak is observed for a wide range of barrier gate settings, where it is not obscured by frequent transmission resonances (red in the color plot). Gate 1 = -1.165 V. Similar traces are obtained when F2 is positive and F3 is used to pinch-off, as well as when Gate 1 is swept. **C**, A pinch-off trace for a different setting of Gate 1 = 4.0 V. The details of conductance are altered. Zero-bias peak is observed only in the high conductance region near F2 = -0.5 V, but not in the lower conductance regions. (Data from N-NW-S Device 1, B = 150 mT, T = 60 mK)

**These traces are typical among other hundreds of barrier gate scans measured in devices 1 and 3: for some gate settings ZBPs are observed, and for other settings ZBP disappears. However, in the present devices it is difficult to separate the effect of tuning the chemical potential from the effect of tuning the barrier transmission. Devices with optimized gate geometries can be used to investigate the Zeeman energy-chemical potential phase diagram of the ZBP.**

## Figure S10: Field dependence from Device 3

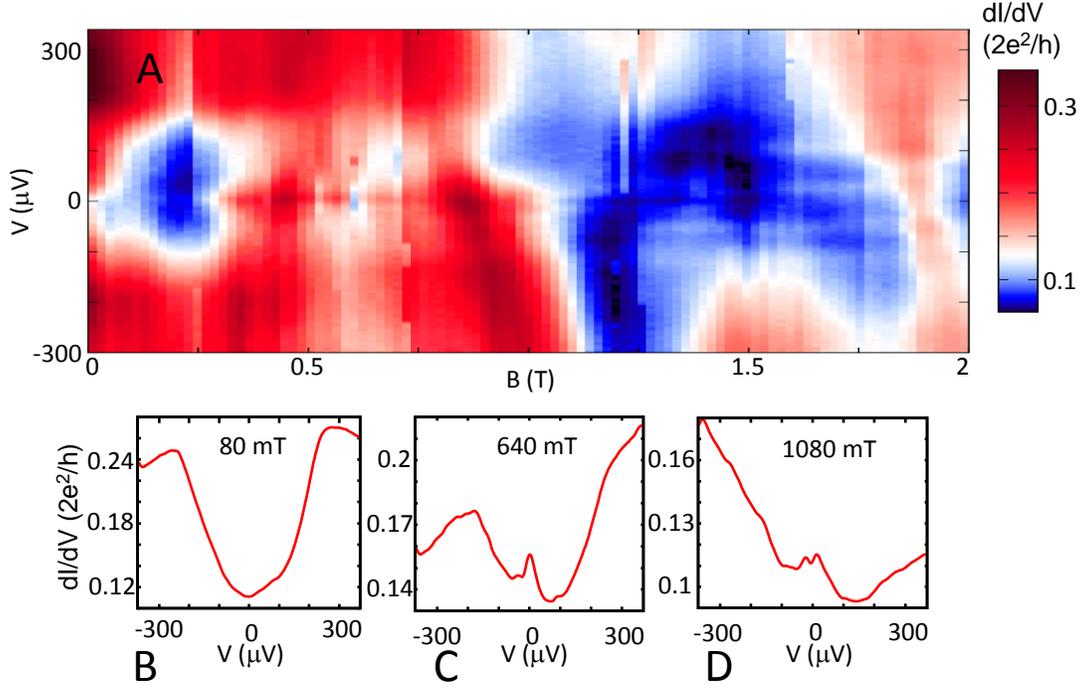

**Field dependences of the ZBP for N-NW-S device 3 are in qualitative agreement with those from devices 1 and 2. A**, Magnetic field map of conductance shows a zero-bias peak that onsets at finite field (0.2 T) and extends to 1 Tesla. Beyond B = 1 T several field-independent resonances are visible. In addition, resonances that are field tunable cross zero bias at several magnetic fields. Local charge rearrangement results in abrupt conductance switches seen in the data. Such "charge noise" is more dramatic in device 3 compared to devices 1 and 2. **B-D**, Linecuts from panel **A**.

All key findings illustrated by data from device 1 throughout the paper and supplementary material are reproduced in device 3. Specifically, we find that ZBP persists in a significant gate range for all gates from the tunnel barrier to gate 3 (the farthest gate from the tunnel barrier for this device). The peak height and width are found to be the same as in device 1 at the lowest temperature (60 mK), temperature dependence was not studied for device 3. The induced gap is of the same magnitude (250 μV). Bound states crossing zero bias are also observed in device 3. Devices 1 and 3 are measured in magnetic field of fixed orientation, therefore comparison of the ZBP height with the spin-orbit field direction is only carried out for device 2.

Figure S11: Gate dependences from Device 3

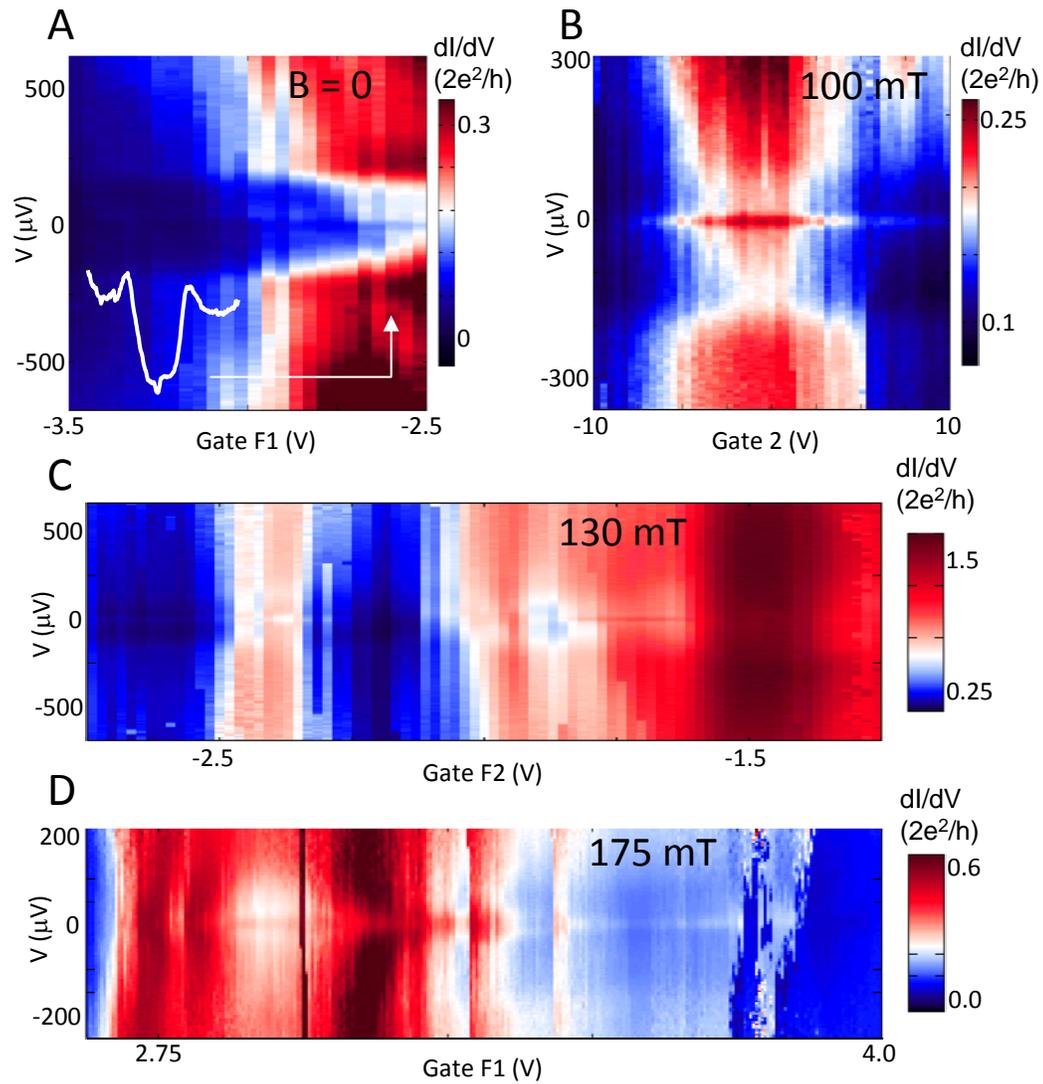

**Gate dependences of the ZBP for N-NW-S device 3 are in qualitative agreement with those from device 1. A**, Zero-field scan of gate F1. A linecut shows the induced gap at a gate setting marked by an arrow. **B**, A scan of gate 2 at finite magnetic field. ZBP is observed in the entire range, an ABS resonance passes through zero bias near zero gate voltage. **C**, Gate F2 scan at finite magnetic field. Regions of ZBP, split peak and absent peak are observed. **D**, Gate F1 scan at finite magnetic field.

## Figure S12: Superconductor-Nanowire-Superconductor devices

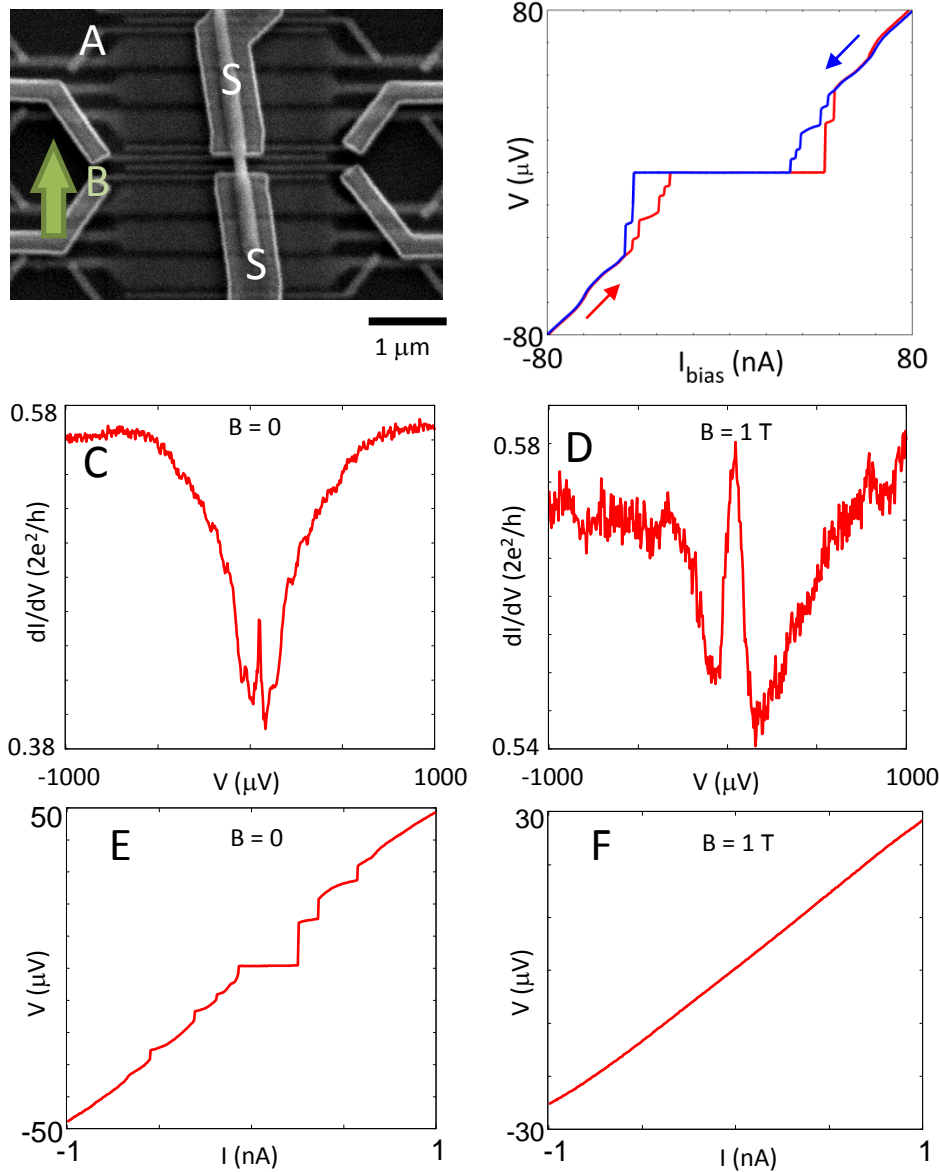

**A**, Indium antimonide nanowires are contacted by two superconducting NbTiN electrodes. **B**, In S-NW-S devices supercurrents are observed (see also Nilsson *et al. (37)*). We found supercurrents exhibiting gate voltage dependence, indicating that superconductivity is induced in the nanowire and confirming proximity effect. **C** and **D**, Zero bias conductance peaks are also observed in S-NW-S devices in voltage-bias experiments. **E** and **F**, Current-biased measurement for the same settings as in panels **C** and **D**. The ZBP in panel **C** is clearly attributed to supercurrent, while a peak at B = 1 T in panel **D** may have the same origin as ZBP in N-NW-S devices. However, in other S-NW-S devices we observe supercurrents extending to B = 1 T. We also observe supercurrents extending to 100's of mT when the superconductor contact spacing is increased to 600 nm. Small supercurrents do not show up as steps in the I-V curves but could still result in enhancement of dI/dV near zero voltage bias in voltage-biased experiments. **This underscores the importance of using a normal metal contact as a tunnel probe in order to exclude supercurrent as an explanation for the zero-bias peak.**

## Figure S13: Normal Metal-Nanowire-Normal Metal device

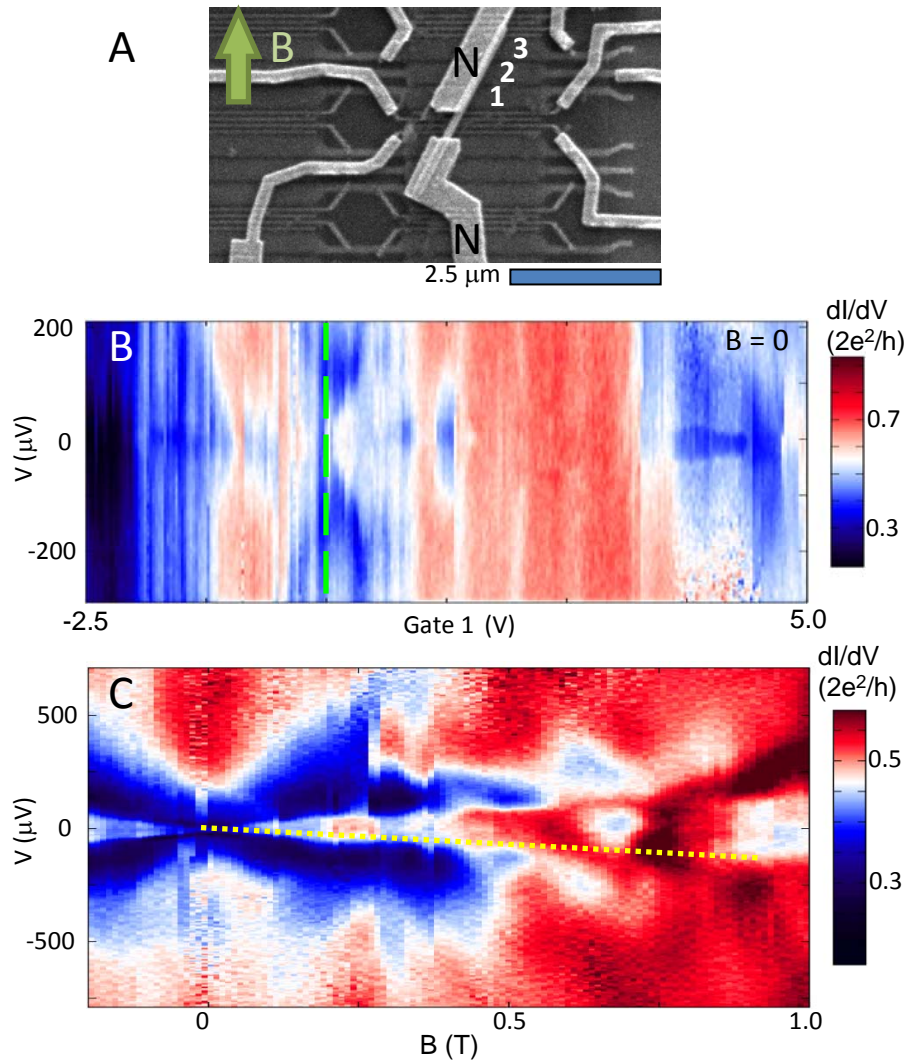

In order to test superconductivity as a necessary ingredient for the rigid ZBP, we have fabricated an N-NW-N device with two Ti/Au contacts to the InSb nanowire (SEM photo in (A)). Wide gates underneath the upper normal contact are tuned, the arrow indicates the direction of the applied magnetic field. **B**, typical scan of Gate 1 near the edge of the half-covering (upper) normal contact. At zero magnetic field no induced gap is observed. A suppressed conductance near zero bias for some gate ranges is not accompanied by quasi-particle peaks characteristic of superconducting gaps. At the green dashed line we observe a zero bias peak (although difficult to see on this scale). In C we investigate the B-field dependence of this zero bias peak and observe that the peak splits. Dashed line is a guide to the eye for the splitting. T = 20 mK for these data.

Figure S14: Normal Metal-Nanowire-Normal Metal device

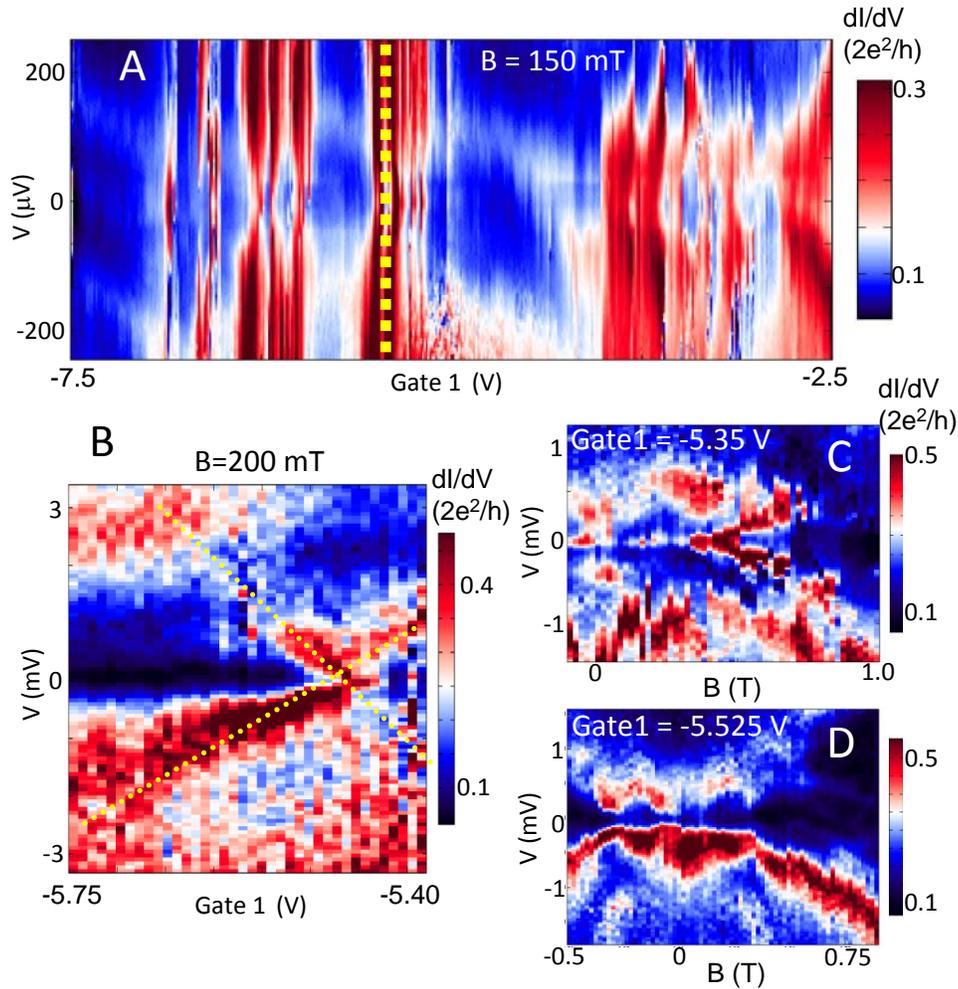

**A**, Data from the same N-NW-N device as in S12: Gate 1 scan at finite magnetic field. The gate range is virtually always clear of zero-bias peaks. However in a small range in gate space of order 100 mV (dotted line) we observe a zero bias peak. (Charge switches in this scan result in the apparent doubling, and sometimes quadrupling, of red resonances that pass through zero.) In **B** we zoom in on this peak in gate range. We observe that the zero bias peak is a near crossing of two gate-tunable resonances (dotted lines). **C** and **D**, Magnetic field dependences obtained for gate settings just left and right of the crossing in panel **B**. When Gate 1 is set just right of the crossing, a zero-bias peak is observed starting at zero magnetic field and splitting at higher field. When Gate 1 is set left of the crossing, a pair of split resonances is observed at higher bias. These resonances continue to split as the field is increased. We conclude that the zero-bias peak that we observe here only occurs in a narrow gate range and is connected to the crossing of two resonant levels. The crossing shifts its position in gate space by a small amount when magnetic field is increased. This produces a zero-bias peak that persists in magnetic field for a few hundred milliTesla. This effect is distinctly different from the rigid ZBP observed in N-NW-S devices, where the peaks persist in BOTH magnetic field and gate voltages for all gates.